\documentclass[twocolumn,amsmath,amssymb,prl]{revtex4}

\usepackage{graphicx}
\usepackage{dcolumn}
\usepackage{bm}

\begin{document}

\title{Tight-binding modeling and low-energy behavior \\
of the semi-Dirac point}

\author{S. Banerjee$^1$, R. R. P. Singh$^1$, V. Pardo$^{1,2}$, and W. E. Pickett$^1$}
\email{wepickett@ucdavis.edu}
\affiliation{$^1$Department of Physics,
 University of California, Davis, CA 95616}
\affiliation{
$^2$Departamento de F\'{\i}sica Aplicada, Universidad
de Santiago de Compostela, E-15782 Santiago de Compostela,
Spain
}




\date{\today}

\begin{abstract}
We develop a tight-binding model description of semi-Dirac electronic spectra,
with highly anisotropic dispersion around point Fermi surfaces,
recently discovered in electronic structure calculations of
VO$_2$/TiO$_2$ nano-heterostructures. We contrast their spectral properties with
the well known Dirac points on the honeycomb lattice relevant to graphene
layers
and the spectra of bands touching each other in
zero-gap semiconductors. We also consider the lowest order dispersion
around one of the semi-Dirac points and calculate the resulting electronic energy levels in an
external magnetic field. We find that these systems support apparently similar electronic
structures but diverse low-energy physics.
\end{abstract}
\maketitle


The ability to prepare graphene (single graphite sheets) \cite{graphene_nat05} has spurred the
study of electronic behavior of this unique system, in which a pair of Dirac
points occur at the edge of the Brillouin Zone \cite{antonio}.
Bands extend linearly (referred to as ``massless Dirac'') both to
lower and higher energy from point Fermi surfaces. This unusual behavior requires special symmetry
and non-bonding bands. In bilayer graphene \cite{graphene_bilayer}, the linearly dispersing bands become quadratic, while
retaining many of the symmetry properties. Spin-orbit coupling in these systems should lead to
a gapped insulator in the bulk. Gapless modes remain at the edge of the system, protected by topological properties
and time-reversal symmetry.
This new state of matter, insulating in the bulk and metallic at the edges,
has been called a topological insulator \cite{qshe_graphene,fu}.

Point Fermi surfaces also arise in gapless semiconductors in which the bands extend quadratically (``massively'')
from a single point separating valence and conduction bands \cite{Tsidilkovski}.
However, these systems are, generically, not topological insulators.
It has been argued that in HgTe quantum wells, where $s$ and $p$ bands overlap each other
at the $\Gamma$ point as a function of well thickness, Dirac-like spectra can also  arise
with exotic topological properties. Due to the enhanced
spin-orbit coupling in these materials,
a state of matter exhibiting quantum spin Hall effect, has been predicted \cite{bernevig} and observed \cite{konig}.

Recent developments in the synthesis of controlled nanostructures, heterojunctions and
interfaces of transition metal oxides represent one of the most promising areas of research in
materials physics.
While several recent studies of oxide interfaces have focused on the polarity discontinuity
that can give rise to unexpected states between
insulating bulk oxides, including conductivity \cite{ohtomo, hwang}, magnetism \cite{magn_IF}, orbital order \cite{pentcheva}, even
superconductivity \cite{IF_sc}, unanticipated behavior unrelated
to polarity can also arise.
The VO$_2$/TiO$_2$ interface involves no polar discontinuity, but only an open-shell charge and local magnetic discontinuity,
according to the change $d^1 \leftrightarrow d^0$ across the interface.

It was recently discovered \cite{vo2_tio2} that a three unit cell slab of VO$_2$ confined within insulating TiO$_2$ possesses
a unique band structure.  It shows four symmetry related
point Fermi surfaces along the (1,1) directions in the 2D Brillouin zone, in this
respect appearing to be an analog to graphene.  The dispersion away from this point is
however different and unanticipated: a gap opens linearly along the symmetry line, but
opens quadratically along the perpendicular direction.  The descriptive picture is that
the associated (electron or hole) quasiparticles are relativistic along the diagonal with an
associated ``speed of light" $v_F$, as they
are in graphene in both directions, but they are non-relativistic in the perpendicular
direction, with an effective mass $m$.  Seemingly the laws of physics (energy vs. momentum) are different along the
two principal axes.  The situation is neither conventional zero-gap semiconductor-like,
nor graphene-like, but has in some sense aspects of both. This kind of spectra was found to be robust
under modest changes in the structure.

Here, we develop a tight-binding model description of this semi-Dirac spectra. We find
that a three-band model is needed, which can be downfolded to two bands at low energies. A variant
of the model, with only two bands, gives rise to anisotropic Dirac spectra, where
one has linearly dispersing modes around point Fermi surfaces, with very different
``speed of light'' along two perpendicular axes.
A common feature of the various systems discussed above are point Fermi-surfaces. From a device
point of view, for example in thinking of p-n or p-n-p junctions, these systems may
share common qualitative features. The actual dispersion, which would give rise to different density of states,
may control more quantitative differences. However, a more fundamental difference may be
in their topological properties.\cite{fu,thouless,haldane,moore}

We begin with a 3-band tight-binding model of spinless fermions [corresponding
to the half-metallic VO$_2$ trilayer although the system could also be nonmagnetic (spin degenerate)],
on a square-lattice, defined by the Hamiltonian
\begin{eqnarray}
{\cal H}=&\sum_{\alpha=1}^3 (\sum_{i}\epsilon_\alpha n_{i,\alpha}+
\sum_{<i,j>} t_\alpha (c_{i,\alpha}^\dagger c_{j,\alpha}+ h.c.)) \\ \nonumber
         &+\lambda_1 \sum_{<i>,\pm} (c_{i,1}^\dagger c_{i\pm \hat x,3} -c_{i,1}^\dagger c_{i\pm \hat y,3} + h.c.)\\ \nonumber
         &+\lambda_2 \sum_{<i>,\pm} (c_{i,2}^\dagger c_{i\pm \hat x,3} -c_{i,2}^\dagger c_{i\pm \hat y,3} + h.c.)\\ \nonumber
\label{eq:tb-realspace}
\end{eqnarray}
with $\epsilon_3>>\epsilon_1,\epsilon_2$, so that we have two overlapping bands $1$ and $2$, with no coupling
between them. Instead, they couple through the third band, by a coupling which changes sign under rotation by
$90$ degrees. Such a coupling can be shown to arise by symmetry between $d$ and $s$ orbitals, for example.
The important aspect is that the coupling vanishes along the symmetry line, allowing the bands to cross
(they have different symmetries along the (1,1) line).  Now, since
the third band is far from the Fermi energy it can be taken as dispersionless. Furthermore, without affecting
any essential physics,
we take $t_1=-t_2=t$ and $\lambda_1=\lambda_2=t'$. Thus, in momentum space the Hamiltonian becomes a $3\times 3$
matrix:
\begin{eqnarray}
H =
\begin{pmatrix}
\widetilde{\varepsilon}_{1k}                & 0                              &         V_k \\
0                                         & \widetilde{\varepsilon}_{2k}      &         V_k \\
V_k                                         & V_k                              & \varepsilon_3
\end{pmatrix}
\label{eq:tightBindingH}
\end{eqnarray}
where the dispersions and coupling are given by
\begin{eqnarray}
\begin{tabular} {r c l}
\(\widetilde{\varepsilon}_{1k} \) & \( = \) & \( \varepsilon_1 + 2t(\cos k_x + \cos k_y) \) \\ \nonumber
\(\widetilde{\varepsilon}_{2k} \) & \( = \) & \( \varepsilon_2 - 2t(\cos k_x + \cos k_y) \) \\
 \(V_k\)                         & \( = \) & \( 2t'(\cos k_x - \cos k_y) \) \nonumber
\end{tabular}
\label{eq:eev}
\end{eqnarray}

Using the fact that orbital 3 is distant in energy, the three-orbital problem can be downfolded to a
renormalized two orbital problem which becomes (neglecting a parallel shift of the two remaining
bands)

\begin{eqnarray}
H =
\begin{pmatrix}
\widetilde{\varepsilon}_{1k}              &     \frac{V_k^2}{\varepsilon_3} \\
\frac{V_k^2}{\varepsilon_3}                                    & \widetilde{\varepsilon}_{2k} \end{pmatrix}
\label{eq:tightBindingH2}
\end{eqnarray}

\begin{figure}[ht]
\begin{center}
\includegraphics[width=\columnwidth,draft=false]{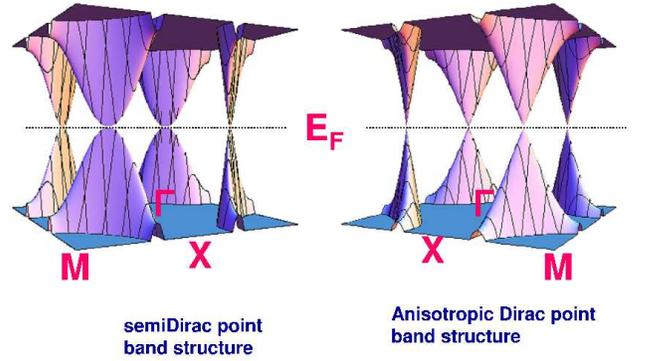}
\caption{On the left, the plot shows the low energy band eigenvalues $E_{q\pm}$
in a region near $E_F$ for the semi-Dirac point.
On the right is the same plot for the anisotropic Dirac point.}
\label{2DE_sD}
\end{center}
\end{figure}

The eigenvalues $E_{k\pm}$ of $H$ as given by \begin{eqnarray}
E_{k\pm}  = \frac{\widetilde{\varepsilon}_{1k}+\widetilde{\varepsilon}_{2k}}{2}
 \pm \frac{1}{2}\sqrt{(\widetilde{\varepsilon}_{1k}-\widetilde{\varepsilon}_{2k})^2
                + 4[{\frac{V_k^2}{\varepsilon_3}}]^2}
\label{eq:eigenValue}
\end{eqnarray}
With some (not very stringent) restrictions on $\varepsilon_1 - \varepsilon_2$ to ensure
that the uncoupled bands actually overlap,
the two bands touch only at the point $\vec k_{sd}$ along the (1,1) lines where
                           $\widetilde{\varepsilon}_{1k}=
                            \widetilde{\varepsilon}_{2k}$,
otherwise the two bands lie on either side of the touching point
(the Fermi energy).
When the 2$\times$2 Hamiltonian is expanded around the semi-Dirac point 
$\vec k_{sd}$ it becomes
\begin{eqnarray}
H =
\begin{pmatrix}
\widetilde{\varepsilon}_{1k}              &  \frac{V_k^2}{\varepsilon_3} \\
\frac{V_k^2}{\varepsilon_3}              & \widetilde{\varepsilon}_{2k}
\end{pmatrix}  \rightarrow \begin{pmatrix}
v_F q_2              &  q_1^2/2m \\
q_1^2/2m             & -v_F q_2 \end{pmatrix}
\label{eq:tightBindingH2b}
\end{eqnarray}
where $q_2$ and $q_1$ denote the distance from $\vec k_{sd}$ along the (1,1) symmetry direction, and the orthogonal (1,${\bar 1}$), respectively. The Fermi velocity $v_F$ and
effective mass $m$ can be related explicitly to the tight binding model parameters, and also calculated by standard ab initio techniques.
The dispersion relation is that found for the three layer slab of VO$_2$
trilayer in
TiO$_2$ at low energy,
\begin{eqnarray}
E_{q\pm} \rightarrow \pm \sqrt{(q_1^2/2m)^2 + (v_F q_2)^2}. \end{eqnarray}
For comparison, the graphene dispersion relation is $E^g_{q\pm}=\pm v_F \sqrt{q_1^2 + q_2^2}$.
A plot of the low-energy dispersion of the model giving rise to a semi-Dirac point in the 2D Brillouin Zone is shown in
Fig. \ref{2DE_sD}.  For the VO$_2$ trilayer,\cite{vo2_tio2} this dispersion holds up to 10-30 meV in the
valence and conduction bands.

A few observations can be made at this point.  First, if the original bands 1 and 2 were
simply coupled by the same anisotropic mixing $V_k$ (without any third band in the picture),
then anisotropic Dirac points (rather than semi-Dirac points) occur along the (1,1) directions.
This dispersion is also shown in Fig. \ref{2DE_sD}.
This type of two-band situation should not be particularly unusual, hence Dirac points in
2D systems are probably not as unusual as supposed, i.e. they are not restricted to
graphene nor are they restricted to high symmetry points.

\begin{figure}[ht]
\begin{center}
\includegraphics[width=6cm,draft=false]{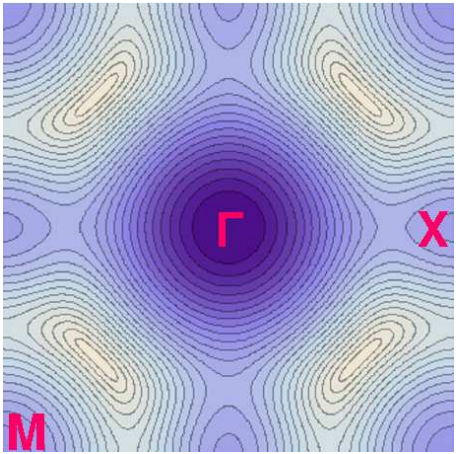}
\caption{Plot showing the Fermi surfaces for electron doping that derive from the low energy excitation spectra of the semi-Dirac point at the $\Gamma-M$ direction in the k$_x$-k$_y$ plane of the square Brillouin zone.}
\label{contours}
\end{center}
\end{figure}

While the constant energy surfaces of our model may appear to be elliptical (the common
situation; the Dirac point has circular FSs), they are actually quite distinct.  As $E\rightarrow$0
the velocity is constant in one direction and is $\sqrt{2mE}$ in the other; the FSs vanish
as needles with their long axis perpendicular to the (1,1) direction. This can be seen in Fig. \ref{contours}, showing the 
constant energy surfaces for electron doping according to our model, showing the 4 semi-Dirac points in the tetragonal k$_x$-k$_y$ Brillouin zone.  The density of
states (DOS) n(E), which
is constant for effective mass systems and goes as $|E|$ for graphene, is proportional to
$\sqrt{|E|}$ at a semi-Dirac point.  When doped, the density of carriers will follow $n(E_F) \propto |E_F|^{3/2}$ behavior.

Another observation is that the {\it same bands} $E_{q\pm}$ can be obtained
from related but distinct low-energy models, such as
\begin{eqnarray}
H_2 =
\begin{pmatrix}
v_F q_2              & iq_1^2/2m \\
-i q_1^2/2m             & -v_F q_2 \end{pmatrix}
\label{eq:tightBindingH2c}
\end{eqnarray}
and
\begin{eqnarray}
H_3 =
\begin{pmatrix}
0              &  q_1^2/2m +i v_F q_2\\
q_1^2/2m -i v_F q_2     & 0 \end{pmatrix}.
\label{eq:tightBindingH2d}
\end{eqnarray}
Although the bands resulting from $H_2$ and $H_3$ are the same, the eigenfunctions
are different and are intrinsically complex for $H_2$ and $H_3$ unlike
for the specific semi-Dirac point we discuss.

One of the issues of most interest to such systems is the behavior in a magnetic field.
Making the usual substitution $\vec q \rightarrow \vec p +\frac{e}{c}\vec A$ with momentum
operator $\vec p$ and vector potential $\vec A$, we find the Landau gauge $\vec A =
B(-x_2,0,0)$ to be the most convenient here.  First, however, we note that the characteristics
of the two directions, the mass $m$ and velocity $v_F$, introduce a natural unit of momentum 
$p_o =m v_F$ and length $x_o = \hbar/p_o$, and of energy $e_o = m v_F^2/2$.
Introducing the atomic unit of magnetic field $B_{\circ}$ such that $\mu_B B_{\circ}$ = 1 Ha, and
the dimensionless field $b = B/B_{\circ}$, units can be scaled away
from the Hamiltonian by defining for each coordinate $x$
\begin{eqnarray}
 x_2  = \Large(\frac{{1}}{\gamma b}\Large)^{2/3} x_{\circ}\tilde{x}_2,
\end{eqnarray}
and similarly for $x_1$.  
Here $\gamma$
is the dimensionless ratio of the two natural energy scales: $\gamma = \mu_B B_{\circ}/(mv_F^2/2)$.
Under this scaling
\begin{eqnarray}
p_1 +\frac{e}{c}A_1 = p_1 -\frac{e}{c}B x_2 \rightarrow p_{\circ}(\gamma b)^{2/3}
 (\tilde{p}_1 - \tilde{x}_2)
\end{eqnarray}
where $\tilde{p}_1, \tilde{x}_1$ are conjugate dimensionless variables, etc.  
Thus all
possible semi-Dirac points (all possible $m$ and $v_F$ combinations) scale to a
{\it single
unique semi-Dirac point}, with the materials parameters determining only the overall energy scale.
There is no limiting case in which the semi-Dirac point becomes
either a Dirac point or a conventional effective mass zero-gap semiconductor.
For the case of trilayer VO$_2$, $\gamma$ does not differ greatly from unity \cite{vo2_tio2}.

Shifting $\tilde{x}_2$ to
$u=\tilde{x}_2 - \tilde{p}_1$,
with conjugate dimensionless momentum $p$, the
Hamiltonian in a field becomes
\begin{eqnarray}
H& = &2 e_o ~(\gamma b)^{2/3} ~[p~ \sigma_z +\frac{1}{2}u^2 ~\sigma_x] \\ \nonumber
 & \equiv &2 e_o~(\gamma b)^{2/3}~h.
\end{eqnarray}
The energy scale is much larger than for conventional orbits though smaller than in
graphene \cite{antonio}, so the VO$_2$ trilayer may display an integer quantum Hall effect
at elevated temperature
as does graphene \cite{roomtempQHE}.

A scalar equation for the eigenvalues can be obtained from $h^2$.  Introducing the
operator $Q = p + i u^2/2$, the eigenvalues of $h^2$ are $Q^{\dag}Q$ and $QQ^{\dag}$,
giving the mathematical problem
\begin{eqnarray}
Q^{\dag}Q \phi_n(u) \equiv \Large(-\frac{d^2~}{du^2} + \frac{1}{4}u^4 -u\Large)\phi_n(u)
        =\varepsilon_n^2\phi_n(u).
\end{eqnarray}
The equation for $QQ^{\dag}$ has the opposite sign of the linear term, with identical
eigenvalues and eigenfunctions related by inversion.  Note that every eigenfunction of $h$
is also an eigenfunction of $h^2$, and that although the potential is negative in the
interval (0,4$^{1/3}$), the eigenvalues $\varepsilon_n^2$ must be non-negative.

\begin{figure}[htbp]
\centering
\includegraphics[width=\columnwidth,draft=false]{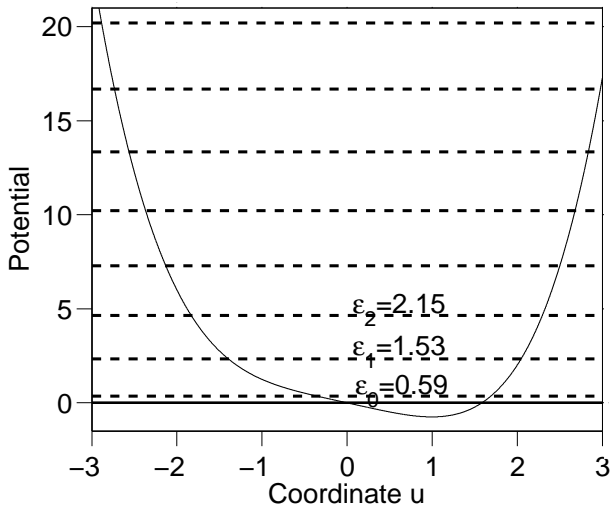}
\caption {Potential energy function for the one-dimensional Schr\"odinger equation and the
resulting quantized energy levels $\varepsilon_n^2$ of $h^2$. The lowest numerical vales
of the three energy eigenvalues
$\varepsilon_n=+\sqrt{\varepsilon_n^2}$ are provided.}\label{fig:quantizedEDia}
\end{figure}

We have obtained the eigenvalues both by precise numerical solution and by WKB approximation,
finding that the latter is an excellent approximation.  Initially neglecting the linear term
in the potential, the WKB condition \cite{wkb}
\begin{eqnarray}
\int_{-\sqrt{2}\epsilon_{n}^{\frac{1}{2}}}^{\sqrt{2}\epsilon_{n}^{\frac{1}{2}}}
         \sqrt{\epsilon^2_{n}-\frac{1}{4}u^4} \ du=(n+\frac{1}{2})\pi
\label{eq:WKBint}
\end{eqnarray}
can be solved to give the WKB eigenvalues for the quartic potential as
\begin{eqnarray}
\epsilon^2_{n}=\Large[3 \sqrt{\frac{\pi}{2}} \frac{\Gamma(\frac{3}{4})}{\Gamma(\frac{1}{4})}\Large]^{4/3}
  (n+\frac{1}{2})^{\frac{4}{3}} = 1.3765 (n+\frac{1}{2})^{\frac{4}{3}}.
\label{eq:WKBEn}
\end{eqnarray}
The linear perturbation corrects the eigenvalues only to second order, which is significant only
for the ground state ($\sim$0.74 versus the numerical solution of 0.59).  The WKB error is less
than 0.01 for the first excited state and gets successively smaller for higher eigenvalues. 
We see then that the semi-Dirac system has eigenvalues in a magnetic field which scale as $B^{2/3}$ and
increase as $(n+\frac{1}{2})^{2/3}$ as $n$ gets large.  Both aspects lie between the behaviors for
conventional Landau levels (linear in $B$, proportional the $n+\frac{1}{2}$) and the Dirac point
behavior (proportional to $\sqrt{B~n}$), as might have been anticipated.
Some low-lying eigenvalues of $h^2$ are
shown in Fig. \ref{fig:quantizedEDia} against the potential well. Note that there is no zero-energy solution as in
the graphene problem.

Another way in which Dirac spectra can arise on a square-lattice can be motivated
in terms of the model of Bernevig {\sl et al.} \cite{bernevig} for HgTe quantum wells.
In their model the two bands crossing each other have $s$ and $p$ characters
respectively. Thus the interband hopping term changes sign under reflection.
This can lead to a ($\sin{k_x} + i \sin{k_y}$) coupling between the bands.
Note that in this model, only a single
Dirac point can occur and it must be at $k=0$, when the
two bands touch each other at that point. In contrast, in the models
discussed here, there are four symmetry related semi-Dirac
(or anisotropic Dirac) points whose location can vary
continuously along the symmetry axis (1,1),
with changes in band parameters.
A feature unique (so far) to the VO$_2$
trilayer system is that point Fermi surface arises in a half metallic
ferromagnetic system where time-reversal symmetry is broken.  Applications
of the VO$_2$ trilayer and related semi-Dirac point systems may provide unusual
spintronics characteristics and applications.

In conclusion, we have developed a tight-binding model description of the semi-Dirac
and anisotropic Dirac spectra relevant to VO$_2$-TiO$_2$ multi-layer systems.
Our tight binding model contains nothing unconventional,
indicating that semi-Dirac and anisotropic point systems are not as rare as has been
assumed.  The low energy characteristics of the semi-Dirac point are intermediate between
those of zero-gap (massive) semiconductors and Dirac (massless) point systems.
The study of such oxide nano-heterostructures has only just begun and they clearly
promise a number of diverse electronic structures and novel phases of matter.


This project was supported by DOE grant DE-FG02-04ER46111 and by
the Predictive Capability for Strongly Correlated Systems team of the Computational
Materials Science Network. V.P. acknowledges financial support from Xunta de Galicia (Human Resources Program).


\end{document}